\documentclass[pra,floatfix,twocolumn,superscriptaddress]{revtex4-1}
\usepackage[final]{graphicx}
\usepackage{times,bbm,amsmath,amssymb}
\usepackage{epsfig,color}
\usepackage{hyperref}
\usepackage{float,siunitx}
\usepackage[caption = false]{subfig}
\usepackage[greek,english]{babel}
\usepackage{thumbpdf,enumerate}
\usepackage{booktabs}
\usepackage{sidecap}
\usepackage[scaled=.8]{couriers}
\usepackage{pstricks}
\usepackage{multirow}
\usepackage{amsmath}
\usepackage{placeins}
\usepackage{pst-grad}
\usepackage{epigraph}
\usepackage{longtable}
\usepackage{booktabs}
\usepackage{gensymb}

\usepackage{pifont}

\newcommand{\ket}[1]{\vert#1\rangle}

\DeclareMathOperator\arctanh{arctanh}

\begin{document}

\title{Quantum simulation of single-qubit thermometry using linear optics}

\author{Luca Mancino}\email{luca.mancino@uniroma3.it}
\affiliation{Dipartimento di Scienze, Universit\`a degli Studi Roma Tre, Via della Vasca Navale 84, 00146, Rome, Italy}

\author{Marco Sbroscia}
\affiliation{Dipartimento di Scienze, Universit\`a degli Studi Roma Tre, Via della Vasca Navale 84, 00146, Rome, Italy}

\author{Ilaria Gianani}
\affiliation{Dipartimento di Scienze, Universit\`a degli Studi Roma Tre, Via della Vasca Navale 84, 00146, Rome, Italy}

\author{Emanuele Roccia}
\affiliation{Dipartimento di Scienze, Universit\`a degli Studi Roma Tre, Via della Vasca Navale 84, 00146, Rome, Italy}


\author{Marco Barbieri} 
\affiliation{Dipartimento di Scienze, Universit\`a degli Studi Roma Tre, Via della Vasca Navale 84, 00146, Rome, Italy}

\begin{abstract}
Standard thermometry employs the thermalisation of a probe with the system of interest. This approach can be extended by incorporating the possibility of using the non-equilibrium states of the probe, and the presence of coherence. Here, we illustrate how these concepts apply to the single-qubit thermometer introduced by Jevtic et al. [S. Jevtic {\it et al.}, Phys.~Rev.~A {\bf 91}, 012331 (2015)] by performing a simulation of the qubit-environment interaction in a linear-optical device. We discuss the role of the coherence, and how this affects the usefulness of non-equilibrium conditions. The origin of the observed behaviour is traced back to the propensity to thermalisation, as captured by the Helmholtz free energy.
\end{abstract}

\maketitle



{\it Introduction.} Thermodynamics provides a description of open systems in terms of the exchange of energy, be it in the form of either heat or work. Although it was developed first in order to give an account of such systems once they have reached the equilibrium with the surrounding environment, it has recently been the object of extensions for treating transient behaviours, irreversibility, and non-equilibrium quantum processes. The knowledge gained through such exertion ranges from fundamental \cite{Jarzynski97,Batalhao15,March16,Rach16}, to more technological issues related to non-equilibrium quantum heat machines \cite{Alicki15,Campisi15,Uzdin15}.

Within such a context, the simplest example considers a single-particle system in contact with a thermal bath; the thermodynamic limit can still be taken, by considering a large collection of identical replicas~\cite{Szilard29}. By isolating a single constituent, the need of accounting for inter-constituent interactions is avoided, and the problem greatly simplified. The attention is then entirely devoted to the internal energy levels of this one constituent, and, if this is a quantum particle, to the coherence among them. Since the presence of quantum coherence underlies the existence of distinctively quantum states, viz. the class of entangled states, it is natural to consider coherence itself as a resource, with appropriate tools for assessing and quantifying its presence~\cite{Baumgratz14,Girolami14,Winter16,Napoli16}. 


These considerations find an immediate application in the context of thermometry, since, on the one hand, we assist at the interaction for a given time of a probe with the monitored system, while, on the other, the probe itself needs being prepared in an informative, hence resourceful, state~\cite{Stace10,Monras11,Brunelli12,Correa15,Guo15}. In~\cite{Jevtic15}, Jevtic {\it et al.} have discussed the implementation of an elementary thermometer with a single qubit: the task is not the estimation of arbitrary temperatures, but the discrimination between two thermal baths at different temperatures. Notably, they have found that limiting the interaction time between the qubit and either bath, thus avoiding thermalisation, result in an improved discrimination. This investigation opens perspectives for realizing temperature measurements at the nanoscale, when the thermometer needs being even smaller than a nano-size thermal bath, e.g. a nanomechanical device~\cite{MAParis} or atomic condensates~\cite{Sabin14,Johnson16,Hohmann16}.

Here we present an experimental investigation of the results of Jevtic {\it et al.} with a linear-optical simulator. We show how one can determine an observable able to discriminate optimally between the two baths, and how the coherence between the two energy level of the qubit influence the performance of the thermometer. Coherence does play a role in the discrimination, but its role is not as simple as a mere enhancement; instead, it affects the time scale at which thermalisation occurs. These features are well captured by the change of the Helmholtz free energy of the probe. Our investigations offer an experimental insight on the roles of quantum non-equilibrium states as probes for thermodynamic processes.

{\it Qubit-bath interaction.} Our thermometer is constituted by a single qubit, governed by its Hamiltonian $\mathbb{H}_S{=}\frac{\hbar \omega}{2}\sigma_z$, where $\sigma_z$ is the $z$-Pauli operator. When isolated, the two levels of the system, the excited state $\vert 0 \rangle$ and the ground state $\vert 1 \rangle$, are separated by $\hbar \omega$, which dictates the energy scale of the protocol. 

\begin{figure}[b]
\includegraphics[width=\columnwidth]{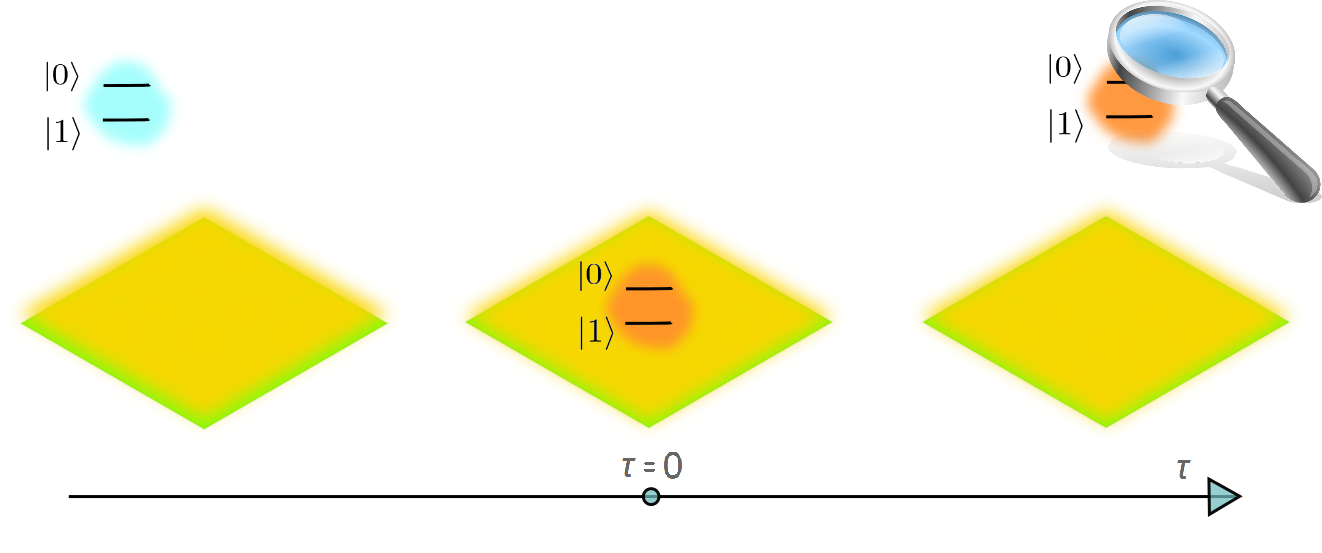}
\caption{Conceptual scheme of the protocol. First, the qubit is initialised in a suitable probe state, then it is put in contact with a thermal bath of unknown temperature, either  $T_1$, or $T_2$. Finally, the qubit is removed from the interaction after a time $\tau$, and measured to infer the working temperature.}
\label{Protocol}
\end{figure}

The interaction of the qubit with a thermal bath, modelled as a gas of non-interacting bosons, results in either of two processes:  $i$) the qubit decays to the ground state transfering its energy to the thermal bath; $ii$) the qubit absorbs an excitation from the reservoir, hence hopping incoherently to the excited state. The rate of the two processes is dictated by the temperature.

Since we are approaching thermodynamics as a theory describing state transformations in the presence of a thermal bath, we introduce a phenomenological model for this interaction as a Generalized Amplitude Damping (GAD) channel \cite{NielsenChuang}. The corresponding map utilizes two couples of Kraus operators. The first one ($E_0$,$E_1$)  describes the decay process ($i$) via a standard Amplitude Damping (AD) channel \cite{NielsenChuang}. The second one ($E_2$,$E_3$) reproduces the inverse process ($ii$); this is an AD too in which the roles of $\vert 0 \rangle$ and $\vert 1 \rangle$ are exchanged (See Appendix). 

The GAD channel is characterized by two parameters: $\gamma$, which represents the decay rate for both the processes and $p$ which is the occurrence probability of the first couple of Kraus operators;  $(1-p)$ is the probability for the other couple. These two parameters are linked to the exact solution of the problem, given by the full Lindblad treatment: $(1-2\bar{N})^{-1}=(1-2p)$ and $(1-\gamma)=\exp[-(1+2\bar{N})\tau]$ where $\bar{N}$ is the average number of excitations in the bath, and $\tau$ is the (dimensionless) interaction time as described in \cite{Jevtic15}. We notice that the Lindblad treatment is only justified in the Markovian limit of the dynamics \cite{Carmichael}. 

{\it Single-qubit thermometry.} Figure [\ref{Protocol}] illustrates our discrimination protocol. At $\tau{=}0^-$ the thermometer is kept isolated and inizialized in the state $\vert \psi \rangle = \cos{\frac{\theta}{2}} \vert 0 \rangle + \sin{\frac{\theta}{2}} \vert 1 \rangle$. At $\tau{=}0$, the qubit is put in contact with the thermal bath which is itself at either a "cold" temperature $T_1$, or a "hot" temperature $T_2{>}T_1$. The different temperatures imply different occupation numbers, $\bar{N_1}$ and $\bar{N_2}$, therefore the qubit undergoes two distinct evolutions depending on the state of the reservoir. Finally, after an interaction time $\tau$, the qubit is isolated again and then measured to determine whether the bath was cold or hot. 

Full thermalization, $\tau {\rightarrow} \infty$, corresponds to the equilibrium regime where the qubit is in a thermal state; conventional thermometry operates within this regime. In our investigation, we extend this analysis to non-equilibrium states. The state of the qubit after the interaction with the reservoir $T_i$ is $\rho_i(\tau)$ ($i=1,2$). The protocol then aims at finding a suitable observable $\hat{\mathbb{G}}(\tau)$ allowing to discriminate $\rho_1(\tau)$ and $\rho_2(\tau)$ optimally \cite{Helstrom76}. The observable $\hat{\mathbb{G}}(\tau)$ is then chosen to maximize the difference $\vert \text{Tr}[\rho_1(\tau) \hat{\mathbb{G}}(\tau)]-\text{Tr}[\rho_2(\tau)\hat{\mathbb{G}}(\tau)] \vert$.

{\it Linear-optics simulation.} We illustrate these concepts by implementing a linear-optics simulator. The main advantage of using simulated dynamics is that it allows to isolate effects stemming genuinely from the process of interest, decoupling all spurious behaviours from other unwanted interactions. The linear-optical approach has demonstrated its ability in replicating sinqle-quantum processes even when conducted in a fully classical regime \cite{Christine,Crespi12,Segev13,Szamait14,Szamait16,Joelle}. Indeed, this takes advantage from the fact that photons are non-interacting particles; using classical light provides a convenient way to obtain a large number of independent replicas. In this work, we adopt this approach for the simulation of an open system, where the qubit is coded in the polarisation, and the coupling to the reservoir occurs via the spatial mode \cite{Almeida07,Cuevas16}. 

\begin{figure}[t!]
\includegraphics[width=\columnwidth]{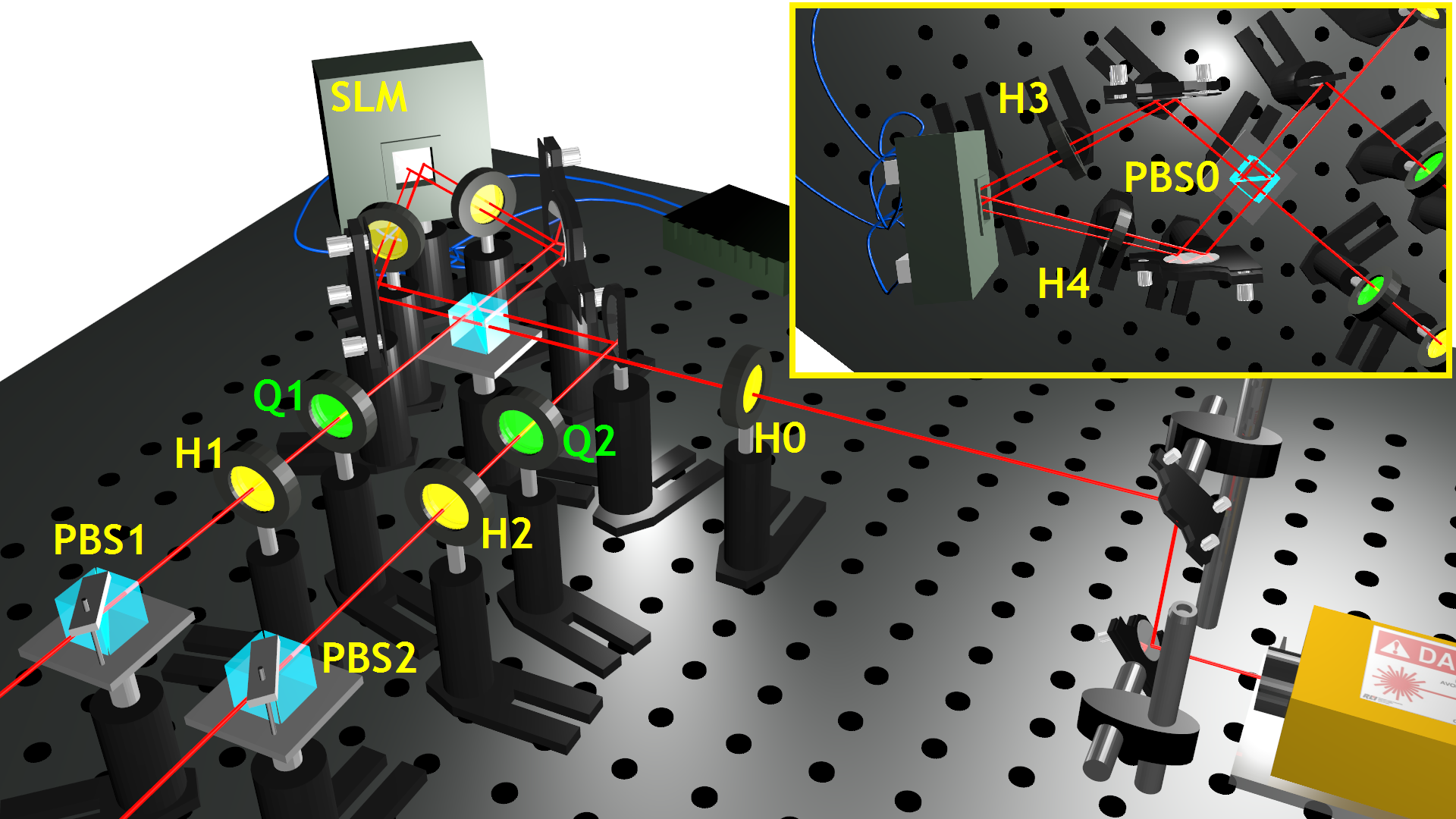}
\caption{Experimental linear-optical simulation. Light is provided by a diode laser emitting $~680\mu$W at 810nm. Its polarisation is controlled by means of the H0 waveplate. The SLM, embedded in a displaced Sagnac interferometer, realises the coupling between polarisation and path, as detailed in the text. At the two outputs, two polarisation analysers, consisting of a quarter wave plate, a half wave plate and a polarising beam splitter are used to characterise the state after the simulated interaction. The two analysis channels, 1 and 2, are kept distinct for practicality, but the results are combined for the analysis. Intensities are detected by a linear diode. Inset: detail of the loops in the Sagnac interferometer. The presence of H3 and H4, both set at an angle of 22.5$^\circ$ makes the polarisation sensitive to the birefringent phase $\phi$ imparted by the SLM. A phase mask is applied, presenting two phase settings: in order to implement $(E_0,E_1)$, the half on the clockwise loop is kept fixed at $\phi{=}0$, while the other half is varied to simulated different interaction times. The mask is then inverted to implement $(E_2,E_3)$.}
\label{SetupThermo}
\end{figure}

Our experimental setup, shown in Fig.\ref{SetupThermo}, consists of a displaced Sagnac interferometer where one of the mirrors is replaced by a spatial light modulator (SLM). By convention, we set the ground (excited) state $\vert 1 \rangle$ ($\vert 0 \rangle$) to be the vertical $\vert V \rangle$ (horizontal $\vert H \rangle$) polarisation state. We initialize the input as a linear polarisation, than send it to the interferometer. The beam is then divided in two using a polarising beam splitter (PBS0) whose outputs constitute the two arms of the Sagnac interferometer. The polarisation is then coupled to the path using two half-wave-plates (H3 and H4) and the SLM that imparts a birefrigent phase $\phi$ (Fig.~\ref{SetupThermo}, inset) \cite{Lemos14}. The mask displayed on the SLM makes sure that such phase is present only on one of the arms while the other arm is unaffected. Overall, this system implements the transformation $\vert H \rangle {\rightarrow} \vert H \rangle$ on the clockwise loop, and $\vert V \rangle {\rightarrow} \left( \cos \frac{\phi}{2} \vert V \rangle + \sin \frac{\phi}{2} \vert H \rangle \right)$ on the counter-clockwise loop. When the two loops are superimposed on PBS0, the horizontal component of the counter-clockwise loop emerges on a separate output; this simulates the incoherent excitation of the qubit corresponding to $E_3$ Kraus operator. The other output is then associated to the complementary event $E_2$. The damping rate is then related to the phase setting as ${\gamma=\sin^2\frac{\phi}{2}}$ (see Appendix). Our device can be programmed to implement the operators $E_0$ and $E_1$ by using a different phase mask on the SLM that now leaves the $\vert V \rangle$ component unaltered. 
\begin{figure*}[t!]
\includegraphics[width=2.1\columnwidth]{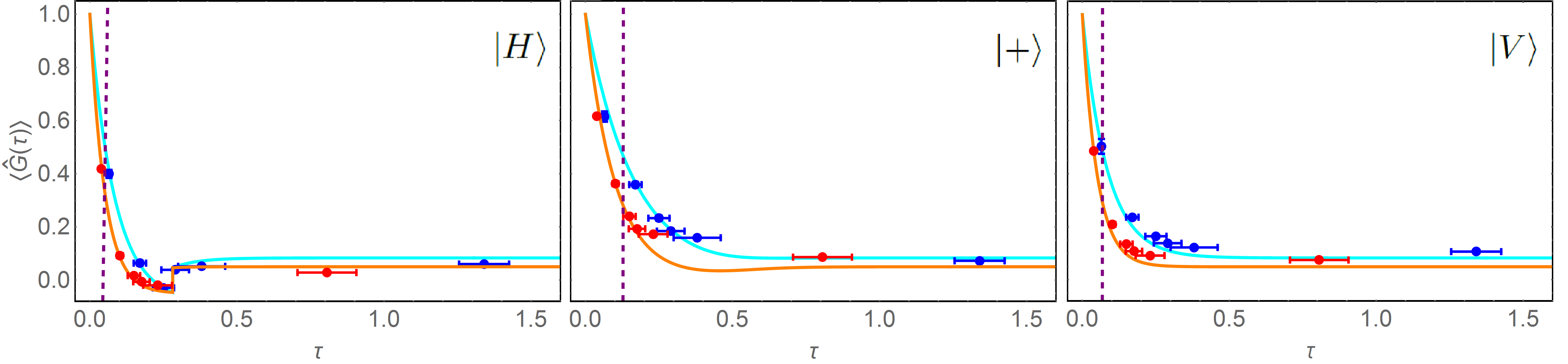}
\caption{Simulated temperature discrimination. The expectation values of  $\hat{\mathbb{G}}(\tau)$ have been inferred from the experimentally reconstructed density matrices, corresponding to three different input states. In the three panels, red dots are for the hot bath $N_2{=}9.5$, blue dots for the cold bath $N_1{=}5.5$; the solid lines show the predicted behaviour. The vertical dashed lines indicate the optimal discrimination time, {\it i.e.} the time for which the difference of the expectation values is maximal. Errors are obtained through a Monte Carlo routine that takes into account the uncertainties on the measured intensities.}
\label{Panel}
\end{figure*}

We reconstruct the density matrix for the qubit after its interaction with the reservoir in the following way: first, we set the interferometer in order to implement the $(E_0,E_1)$ transformation, and perform polarisation tomography~\cite{James01}, without distinguishing the outputs of the interferometer. We repeat the same operation, using the second setting $(E_2,E_3)$. The two experimentally reconstructed matrices are then summed with the opportune weighting $p$, $(1-p)$ to obtain the state after the complete interaction~\cite{note}. We then have access to the state of the qubit at different evolution times and for both the hot and cold baths, corresponding to different choices of the phase $\phi$, and of the weight $p$. 

The results for the discrimination protocol are shown in Fig. \ref{Panel}, where we plot the expectation values of $\hat{\mathbb{G}}(\tau)$ for the two baths associated to three different input states: $\ket{H}$, $\ket{+}{=}(\ket{H}{+}\ket{V})/\sqrt{2}$, and $\ket{V}$. In the three cases, the observed values follow closely the predictions, and demonstrate that $\hat{\mathbb{G}}(\tau)$ serves well the purpose of discriminating between the two possible temperatures. The maximal separation occurs at short times, well before the qubit has reached full thermalisation with the reservoir. These three states are associated to three different strategies: $\ket{V}$ corresponds to the ground state of the qubit, hence we simulate the standard procedure of heating the thermometer; $\ket{H}$ corresponds to the excited state, hence we simulate the cooling of the thermometer; finally, $\ket{+}$ is a coherent strategy, based on the superposition of a hot and a cold thermometer. As expected, in the steady-state regime, the use of any of the three state is equivalent, as thermalisation erases any information on the initial state. Furthermore, the presence of the coherence does not help either in implementing a more effective thermometer, since the optimal separation between $\text{Tr}[\rho_1(\tau) \hat{\mathbb{G}}(\tau)]$ and $\text{Tr}[\rho_2(\tau)\hat{\mathbb{G}}(\tau)]$ weakly depends on the input, nor a faster thermometer, as the optimal measurement time occurs at shorter times for the ground state $\ket{V}$. The main advantage of using the state $\ket{+}$ is in the possibility of maintaining a satisfactory discrimination ability for longer times, as shown by the width of the separation between the two curves; in practical applications, this eases the requirements on the controlled interaction between the qubit and the reservoir.

{\it Free energy and the discrimination power of the single-qubit thermometer.} The origin of this behaviour has been traced back to the different trajectories of the Block vector associated to the qubit in the presence of either bath~\cite{Jevtic15}. Here, we show that this can also been understood in purely thermodynamic terms by looking at the variation of the Helmholtz free energy $\Delta F$ that the qubit undergoes during the interaction process with the external thermal bath. Taking into account the isothermal transformation of the system between its initial ($\rho_{in}$) and final ($\rho_{out}$) states, we can express Helmholtz free energy change as: $\Delta F = \Delta U - T \Delta S$, where $\Delta U=\text{Tr}[\mathbb{H}_S{(\rho_{\rm{in}}-\rho_{\rm{out}})}]$ is the difference in the internal energy, $k_B$ is Boltzmann's constant, and $\Delta S$ is the difference in the Von Neumann entropies, $\Delta S{=}-k_B\text{Tr}[\rho_{\rm{out}}\log(\rho_{\rm{out}})]$, since $S(\rho_{\rm{in}})=0$, being $\rho_{\rm{in}}$ a pure state. 

\begin{figure}[b]
\includegraphics[width=\columnwidth]{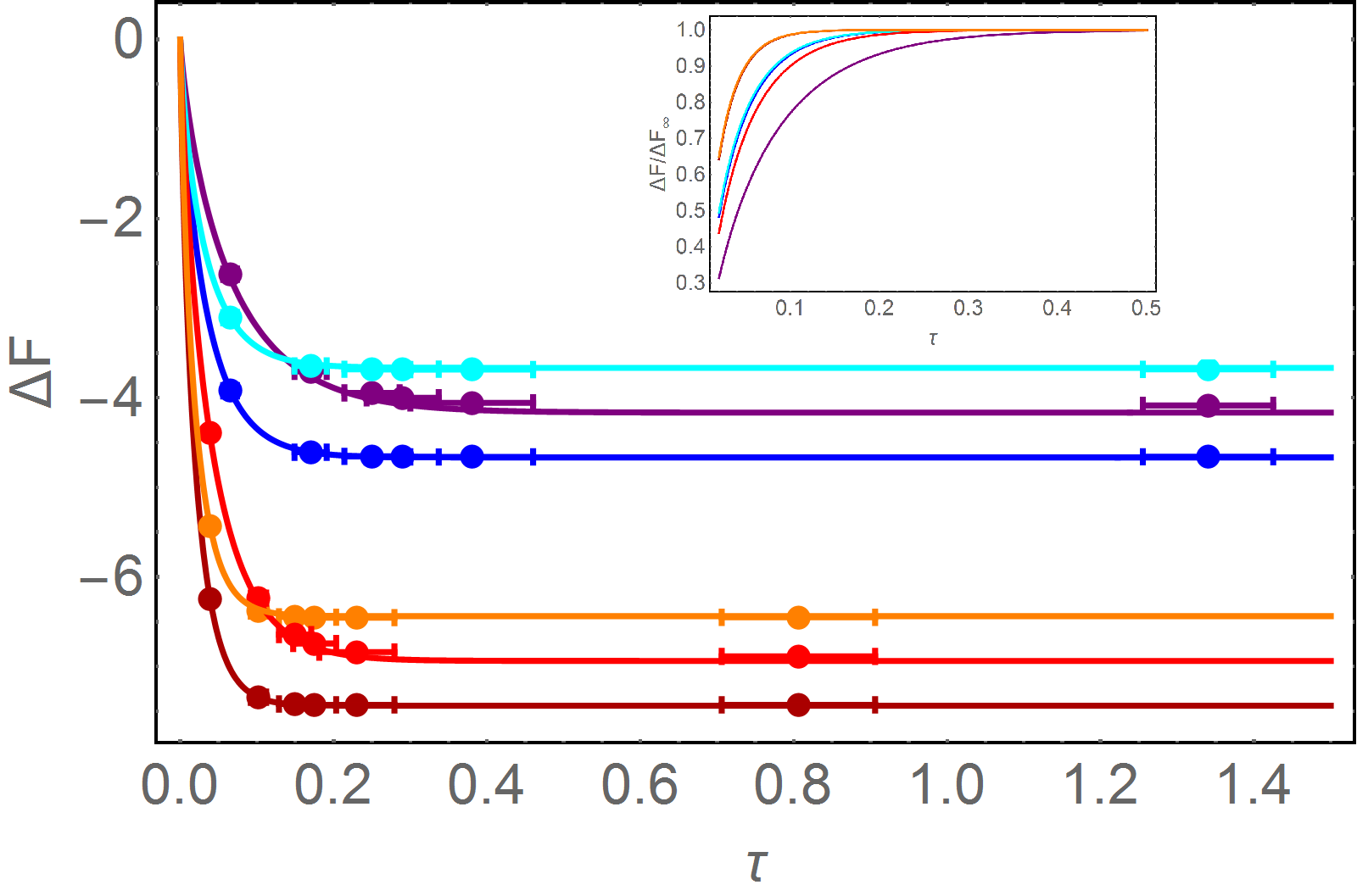}
\caption{Variation of the free energy during the evolution of the system. The points are the free energies of the output states extracted from the experimental density matrices, using different input states: $\vert H \rangle$ (dark red and blue), $\vert D \rangle$ (red and purple) and $\vert V \rangle$ (orange and cyan). The solid curves are the predicted behaviours. The evolution in the presence of the hot (cold) bath, results in a larger (smaller) variation of the free energy. Inset: Predicted variation of the free energy, normalised to its limit value $\Delta F_\infty$ at large times.}
\label{4}
\end{figure}

It can be shown that the Von Neumann entropy of the system increases monotonically, a signature of Markovian dynamic \cite{Chruscinski14}; such a unidirectional information flow between the system and the thermal bath results in a decrease in the Helmholtz free energy, as expected for spontaneous transformations. The observed variation as a function of the time is shown in Fig. \ref{4}, where $\Delta F$ is measured in units of $\hbar\omega$. As expected, the variation is more pronounced when the qubit interacts with the hot bath, and there is a clear dependence of the final value on the initial state, due to the different energy variation $\Delta U$.

Qualitative assessments on the functioning of the thermometer can be inferred by the dynamics of the variation of $\Delta F$, and how this is affected by the coherence in the initial qubit state; this not only fixes the limit value at the thermalisation, but also dictates the {\it speed} at which this occurs. Since optimal discrimination exploits the transient states of the qubit, this constitutes a critical parameter for its performance. In the case of initialisation in the coherent superposition $\ket{+}$, we are able to slow down the thermalisation, and we do so in a different manner for the two possible evolutions. Therefore, we obtain a longer transient that assists the discrimination. The initialisation in the two energy states $\ket{H}$ and $\ket{V}$ results in a similar, shortened time scale, as observed in the curves of Fig.~\ref{Panel}. These behaviours are made more evident when considering the variation of the Helmholtz free energy rescaled to the asymptotic value $\Delta F_\infty$, for all the distinct input states and reservoirs (Fig.~\ref{4}, inset).

{\it Conclusions and perspectives.} We have shown an experimental investigation of the results of Jevtic \textit{et al.} with a linear-optical simulator. Despite the simplicity of the protocol, interesting insights are obtained on the usefulness of non-equilibrium states, and the interplay with the coherence of the system. The capacity of the thermometer in distinguishing between hot and cold thermal bath strongly depends on the initial state of the qubit: while starting from the ground state might allow for a faster operation, coherence allows to maintain a discrimination ability for longer times. These conclusions are supported by the behaviour of the Helmholtz free energy of the system. Within this framework, the availability of a simulation tool, which can be also applied to quantum light, may stimulate explorations to more complex dynamics. This platform could be a testbed for introducing methods of quantum metrology in thermometry~\cite{Correa15,Depasquale}, or ideas from thermometry in the monitoring of quantum channels, establishing connections between thermodynamic potentials and ultimate limits to the precision.  \\

\textit{Note}: During preparation of this manuscript we became aware that similar work was being pursued by W.K. Tham et al \cite{stein}.\\

{\it Acknowledgements.} We are grateful to Antonella De Pasquale for insightful feedback on the manuscript. We would like to thank Fabio Sciarrino for the loan of scientific equipment. MB has been supported by a Rita Levi-Montalcini fellowship of MIUR. 

\section*{Appendix}
{\it Appendix A} We study the interaction of a qubit with a thermal bath according to a standard master equation. The qubit is first prepared in a known pure state with Bloch vector $r(0)=(r_x,r_y,r_z)$. After interaction from the reservoir. and subsequent detachment after a time $\tau$, the Bloch vector is given by:

\begin{equation}
r(\tau,T)=\begin{pmatrix}
r_x e^{-(1+2\bar{N})\tau/2} \\
r_y e^{-(1+2\bar{N})\tau/2} \\
\frac{e^{-(1+2\bar{N})\tau}(1+(1+2\bar{N})r_z)-1}{1+2\bar{N}}
\end{pmatrix}
\label{Blochtrasformation}
\end{equation}
as shown in \cite{Jevtic15}. Here, time is taken as a dimensionless parameter, as the actual time is normalised to the characteristic spontaneous emission of the probe. The resulting dynamics is described by a CP-map corresponding to a generalized amplitude damping (GAD) channel, providing a suitable form for the simulation  of the non-equilibrium dynamic processes, and the thermalisation of the probe. The GAD channel is composed of the following Kraus operators: 

\[
E_0=\sqrt{p}\begin{pmatrix}
1 & 0 \\
0 & \sqrt{1-\gamma}
\end{pmatrix},
\;\;\;\;
E_1=\sqrt{p}\begin{pmatrix}
0 & \sqrt{\gamma} \\
0 & 0
\end{pmatrix}
\]

\[
E_2=\sqrt{1-p}\begin{pmatrix}
\sqrt{1-\gamma} & 0 \\
0 & 1
\end{pmatrix},
\;\;\;\;
E_3=\sqrt{1-p}\begin{pmatrix}
0 & 0 \\
\sqrt♪{\gamma} & 0
\end{pmatrix}
\] 
This corresponds to two independent amplitude damping channels with the same damping rate $\gamma$, acting with a probabilities $p$ and $1{-}p$, working in opposite sense: one damps population in the ground state, the other in the upper state.

As the transformation performed by the channel is: $(r_x,r_y,r_z){\rightarrow}\left( r_x \sqrt{1-\gamma}, r_y \sqrt{1-\gamma}, \gamma (2p-1)+r_z(1-\gamma) \right)$, a satisfactory control of the bath temperature and interaction time theoretical parameters can be realized via the parameters $p$ and $\gamma$. This mapping can be realized performing Bloch vectors component by component equalities, leading to the following results: $(1-2\bar{N})^{-1}=1-2p$ and $1-\gamma=\exp[-(1+2\bar{N})\tau]$.

\begin{figure}[t!]
\includegraphics[width=\columnwidth]{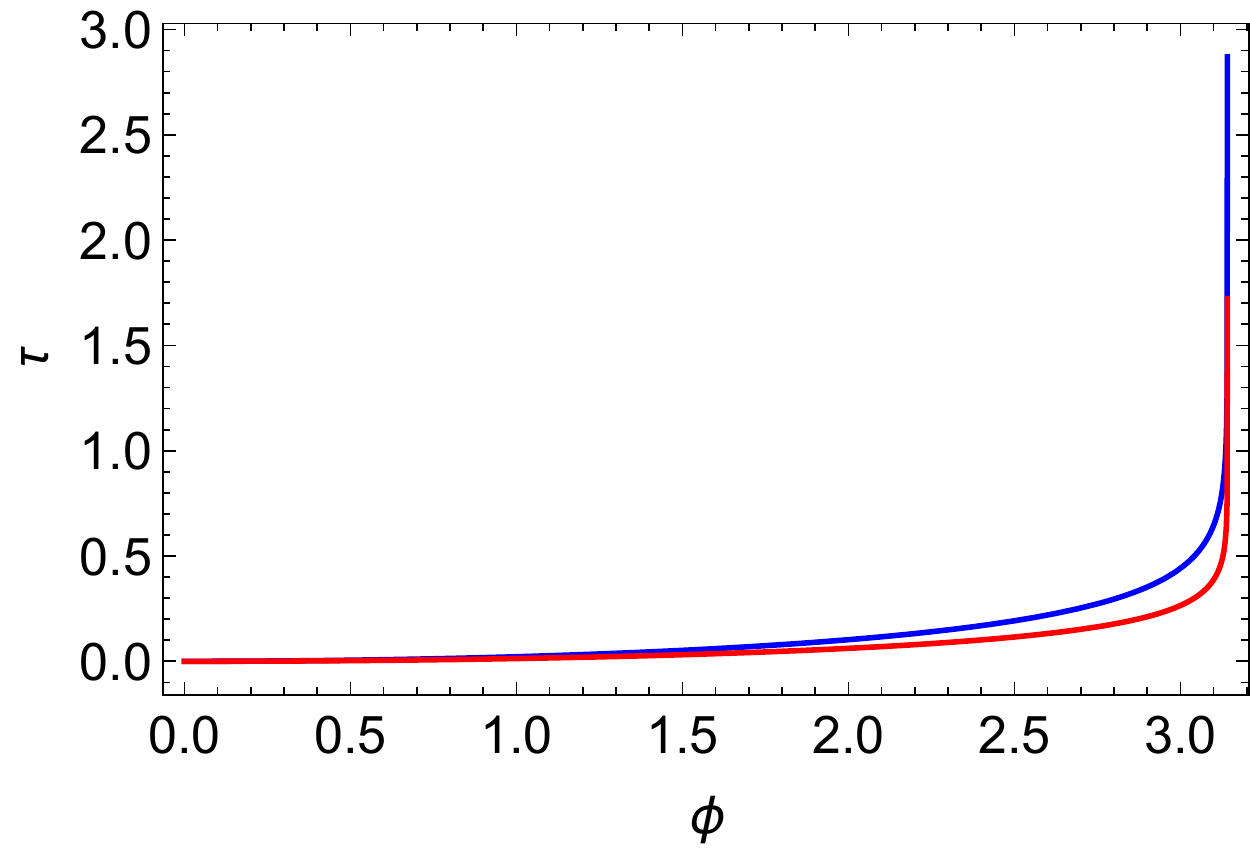}
\caption{Interaction time dispersion as a function of the SLM birefrangent phase $\phi$. Red line is for the hot bath $\hat{N}_2=9.5$ while blue line is for the could bath $\hat{N}_1=5.5$. By using different $\phi$ parameters, the interaction time between the thermometer probe and the reservoir can be opportunely tuned.}
\label{TimeDispersion1220}
\end{figure}

{\it Appendix B.} The experimental implementation of our channel provides is based on the use of a Sagnac interferometer in which one of the mirrors have been substituted by a spatial light modulator (SLM) between two half wave plates (HWPs), set to perform a Hadamard transformation ($\hat{H}_g$) on the polarisation states: $\hat{H}_g\ket{H}=\ket{+}$, $\hat{H}_g\ket{V}=\ket{-}$. In the general case, the input $\ket{\psi}=\alpha\ket{H}+\beta\ket{V}$ is first split on the PBS0, coupling the polarisation to the arm within the Sagnac: $\alpha\ket{H}_{cw}+\beta\ket{V}_{ccw}$, where $cw$ $(ccw)$ indicates the (counter-)clockwise direction in the interferometer.

To reproduce one pair of the GAD Kraus operators, the SLM has been used to implement the unitary transformation $\hat{U}=\mathbb{I}$ on the $cw$ mode, and $\hat{U}=\exp[i\frac{\phi}{2}] \vert H \rangle \langle H \vert + \exp[-i\frac{\phi}{2}] \vert V \rangle \langle V \vert$ for the $ccw$ mode, where $\phi$ represents the birefringent phase imparted by the SLM. The overall transform is then $\hat{H}_g \hat{H}_g = \mathbb{I}$ on $\vert H \rangle$, and $\hat{H}_g \hat{U} \hat{H}_g$ on $\vert V \rangle$; this results in  $\vert H \rangle$ remaining unaltered, while  $\vert V \rangle$ is transformed as $\cos(\phi/2) \vert V \rangle + \sin(\phi/2) \vert H \rangle$. PBS0 then directs photons in the modes $cw$ and $ccw$ towards the two output modes 1 and 2, in a polarisation-dependent fashion; the unnormalised states on the two outputs are then: $\alpha\ket{H}_1+\cos{(\phi/2)}\ket{V}_1$, and simply $\sin{(\phi/2)}\beta\ket{V}_2$. By direct comparison with the theoretical treatment in Appendix A, it is possible establish the relation between $\gamma$ and $\phi$ as $\gamma=\sin^2 \left( \frac{\phi}{2} \right)=1-\exp[-(1+2\bar{N})]\tau$. The phase mask of the SLM can be modified to applying the unitary transformation $\hat{U}=\mathbb{I}$ for the $ccw$ mode, and $\hat{U}=\exp[i\frac{\phi}{2}] \vert H \rangle \langle H \vert + \exp[-i\frac{\phi}{2}] \vert V \rangle \langle V \vert$ for the $cw$ mode; the density matrices coming from the other Kraus operators can thus be obtained.

Finally, the link between the interaction time and the birefringent phase of the SLM can be easily obtained resulting in:
\begin{equation}
\tau=-\frac{\log \left( 1 - \sin^2 \left( \frac{\phi}{2} \right) \right)}{1+2\bar{N}}
\end{equation}
shown in Fig. \ref{TimeDispersion1220}; this has been used as a calibration curve, for use with the experimental values of $\phi$. 

\begin{figure}[t]
\includegraphics[width=\columnwidth]{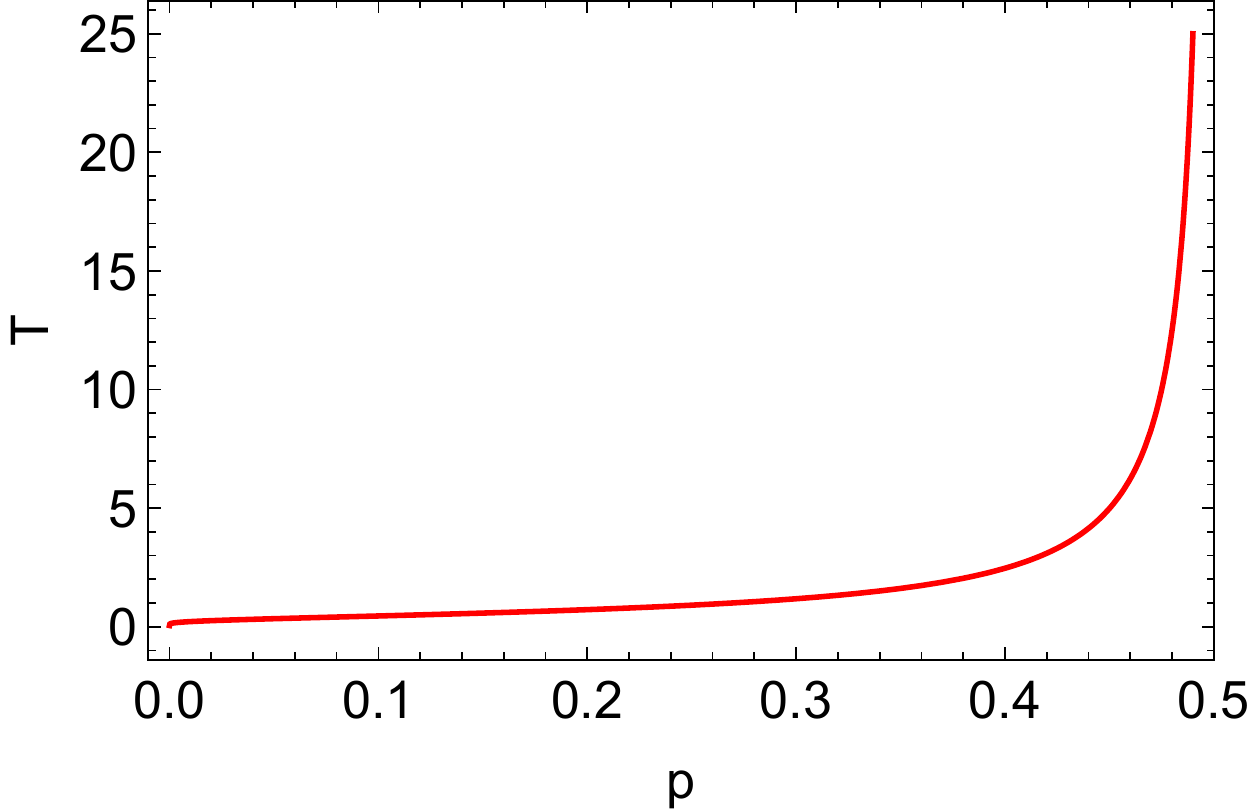}
\caption{Temperature dispersion as a function of the experimental parameter $p$. The cold reservoir temperature $\bar{N}_1=5.5$ corresponds to $p=0.458$ while $\bar{N}_2=9.5$ corresponds to $p=0.475$.}
\label{TempDispersion}
\end{figure}

The other experimental parameter to control is represented by the weighting $p$ which we employ for the data processing. This can be put in relation with the effective temperature $T$ of the bath; this is measured in units of $\hbar \omega/k_B$, in order to get a dimensionless parameter. The relation linking $p$ with $T$ is be obtained by evaluating the boson occupation number $\hat{N}$ as a function of temperature; this leads to the final result:
\begin{equation}
T=\frac{1}{2 \arctanh(1-2p)}
\end{equation}
The relative curve is shown in Fig.\ref{TempDispersion}.


\begin{thebibliography}{99}
\bibitem{Jarzynski97} C. Jarzynski, \prl~{\bf 78}, 2690 (1997).
\bibitem{Batalhao15} T.B. Batalhao {\it et al.}, \prl{\bf 115}, 190601 (2015).
\bibitem{March16} M.A. Garc\'ia-March, T. Fogarty, S. Campbell, T. Busch, and M. Paternostro, {arXiv:1604.03378}, (2016)
\bibitem{Rach16} N. Rach, S. Montangero, and M. Paternostro, {arXiv:1605.07476} (2016)
\bibitem{Alicki15} R. Alicki and D. Gelbwaser-Klimovsky, New J. Phys. {\bf 17}, 115012 (2015).
\bibitem{Campisi15} M. Campisi, J. Pekola, and R. Fazio, New J. Phys. {\bf 17}, 035012 (2015).
\bibitem{Uzdin15} R. Uzdin, A. Levy, and R. Kosloff, Phys. Rev. X~{\bf 5}, 031044 (2015).
\bibitem{Szilard29} L. Szilard, Z. Phys. {\bf 53}, 840 (1929).
\bibitem{Baumgratz14} T. Baumgratz, M. Cramer, and M.B. Plenio, \prl~{\bf 113}, 140401 (2014).
\bibitem{Girolami14} D. Girolami, Phys. Rev. Lett. {\bf 113}, 170401 (2014).
\bibitem{Winter16} A. Winter, and D. Yang \prl~{\bf 116}, 120404 (2016).
\bibitem{Napoli16} C. Napoli et al., \prl~{\bf 116}, 150502 (2016).
\bibitem{Stace10} T.M. Stace, \pra~{\bf 82}, 011611(R) (2010).
\bibitem{Monras11} A. Monras, and F. Illuminati, \pra~{\bf 83}, 012315 (2011).
\bibitem{Brunelli12} M. Brunelli, S. Olivares, M. Paternostro, and M.G.A. Paris, \pra~{\bf 86}, 012125 (2012).
\bibitem{Correa15} L. A. Correa, M. Mehboudi, G. Adesso, and A. Sanpera, \prl~{\bf 114}, 220405 (2015).
\bibitem{Guo15} L.-S. Guo, B.-M. Xu, J. Zou, and B. Shao, \pra~{\bf 92}, 052112 (2015).
\bibitem{Jevtic15} S. Jevtic, D. Newman, T. Rudolph, and T.M. Stace, \pra~{\bf 91}, 012331 (2015).
\bibitem{MAParis} M. Brunelli, S. Olivares and M. G. A. Paris, Phys. Rev. A, \pra~{\bf 84}, 032105 (2011).
\bibitem{Sabin14} C. Sab\'in, A. White, L. Hackermuller, and I. Fuentes, Sci. Rep. {\bf 4}, 6436 (2014). 
\bibitem{Johnson16} T. H. Johnson, F. Cosco, M. T. Mitchison, D. Jaksch, and S. R. Clark, Phys. Rev. A {\bf 93}, 053619 (2016).
\bibitem{Hohmann16} M. Hohmann, F. Kindermann, T. Lausch, D. Mayer, F. Schmidt, and A. Widera, Phys. Rev. A {\bf 93}, 043607 (2016).
\bibitem{NielsenChuang} M. Nielsen, I.L. Chuang, {\it Quantum computation and quantum information}, {Cambridge University Press, Cambridge} (2000)
\bibitem{Carmichael} H. Carmichael, An Open Systems Approach to Quantum Optics, (Springer, 1993)
\bibitem{Helstrom76} C. W. Helstrom, Quantum Detection and Estimation Theory, vol. 123 of Mathematics in Science and Engineering (Academic Press, New York, 1976).
\bibitem{Christine} A. Schreiber, A. Gabris, P.P. Rohde, K. Laiho, M. Stefanak, V. Potocek, C. Hamilton, I. Jex and C. Silberhorn, Science {\bf 336} 55 (2012).
\bibitem{Crespi12} A. Crespi, S. Longhi and R. Osellame, Phys. Rev. Lett. {\bf 108}, 163601 (2012)
\bibitem{Segev13} M. Segev, Y. Silberberg and D.N. Christodoulides, Nat. Photon. {\bf 7}, 197 (2013)
\bibitem{Szamait14} T. Eichelkraut, C. Vetter, A. Perez-Leija, H. Moya-Cess, D. N. Christodoulides, A. Szameit, Optica {\bf 1}, 268 (2014)
\bibitem{Szamait16} D. N. Biggerstaff, R. Heilmann, A. A. Zecevik, M. Gr\"arfe, M. A. Broome, A. Fedrizzi, S. Nolte, A. Szameit, A. G. White, I. Kassal, Nat. Commun. {\bf 7}, 11282 (2016).
\bibitem{Joelle} J. Boutari, A. Feizpour, S. Barz, C. Di Franco, M.S. Kim, W.S. Kolthammer and I.A. Walmsley, arXiv:1607.00891 (2016)
\bibitem{Almeida07} M. P. Almeida {\it et al.}, Science {\bf 316}, 579 (2007).
\bibitem{Cuevas16} A. Cuevas {\it et al.}, {arXiv:1604.08350} (2016).
\bibitem{Lemos14} G. Barreto Lemos et al. , Phys. Rev. A {\bf 89}, 042119 (2014).
\bibitem{James01}  D.F.V. James, P.G. Kwiat, W.J. Munro, and A.G. White, \pra~{\bf 64}, 052312 (2001).
\bibitem{note} We have controlled that this procedure delivers the same results as running the matrix reconstruction algorithm on weighted data. The main advantage of our strategy consists in introducing an extra reconstruction step, serving as a further control test.
\bibitem{Chruscinski14} D. Chru\'sci\'nski and S. Maniscalco, \prl{\bf 112}, 120404 (2014).
\bibitem{Depasquale} A. De Pasquale, D. Rossini, R. Fazio and V. Giovannetti, {arXiv:1504.07787} (2015)
\bibitem{stein} W.K. Tham, H. Ferretti, A.V. Sadashivan, A.M. Steinberg, \textit{to be published} (2016)
\end{thebibliography}
\end{document}